\documentclass[prl,aps,twocolumn,superscriptaddress]{revtex4-2}

\usepackage[dvips]{graphicx}
\usepackage{amssymb,amsfonts,amsmath,gensymb}
\usepackage{color}
\usepackage[normalem]{ulem}
\usepackage{natbib}
\usepackage[hidelinks]{hyperref}
\usepackage{cleveref}

\renewcommand{\vec}[1]{\boldsymbol{#1}}

\newcommand{\f}{\ensuremath{\vec{f}_{d}}}
\renewcommand{\r}{\ensuremath{\vec{r}}}

\newcommand{\qi}{\ensuremath{\vec{q}_{i}}}
\newcommand{\mat}[1]{\ensuremath{\boldsymbol{#1}}}

\begin{document}

\title{Colloidal transport in twisted lattices of optical tweezers}

\author{Nico C.\ X.\ Stuhlm\"uller}
\affiliation{Theoretische Physik II, Physikalisches Institut, Universit{\"a}t Bayreuth, D-95440 Bayreuth, Germany}

\author{Thomas M.\ Fischer}
\affiliation{Experimatalphysik X, Physikalisches Institut, Universit{\"a}t Bayreuth, D-95440 Bayreuth, Germany}

\author{Daniel de las Heras}
\affiliation{Theoretische Physik II, Physikalisches Institut, Universit{\"a}t Bayreuth, D-95440 Bayreuth, Germany}

\begin{abstract}
We simulate the transport of colloidal particles driven by a static and homogeneous drift force, and subject to the optical potential created by two lattices of optical tweezers.
The lattices of optical tweezers are parallel to each other, shifted, and rotated by a twist angle.
Due to a negative interference between the potential of the  two lattices, flat channels appear in the total optical potential.
At specific twist angles, known as magic-angles, the flat channels percolate the entire system and the colloidal particles can then be transported using a weak external drift force.
We characterize the transport in both square and hexagonal lattices of twisted optical tweezers.
\end{abstract}

\maketitle

\section{Introduction}

Optical tweezers~\cite{Ashkin1986} use optical gradient forces to manipulate micrometer-sized colloidal particles.
Lattices of optical tweezers arranged in arbitrary patterns can be created e.g.\ using diffractive optical elements~\cite{doi:10.1063/1.1148883}, combining beam splitters and refractive optics~\cite{Fallman:97},
by means of computer-generated holograms~\cite{LIESENER200077}, and even rapidly moving a single beam among different locations such that the desired pattern emerges as a results of a time-averaged optical potential~\cite{Sasaki:91,577338,Grier2003}.

Periodic lattices of optical tweezers in combination with a driving force are widely used to sort particles~\cite{MacDonald2003,PhysRevLett.89.128301,Lindenberg,https://doi.org/10.1002/elps.200800484}.
Using a three-dimensional periodic optical lattice, MacDonald et al~\cite{MacDonald2003} were able to sort particles exploiting the differences in the interactions between the particles and the optical lattice.
Also, as shown by Lacasta et al~\cite{Lindenberg}, particles moving in a periodic optical potential can behave differently according to their size or particle index of refraction.

Motivated by the emerging field of twistronics~\cite{PhysRevB.95.075420}, we adapt here the setup of Lacasta et\ al.~\cite{Lindenberg} to model two sets of periodic lattices of optical tweezers that are parallel to each other and are also twisted by a given twist angle.
Using computer simulations we study the transport of colloidal particles subject to the combined potential of both lattices and driven by a uniform and time-independent drift force.
At specific twist angles, known as magic-angles, the transport is more efficient due to the formation of flat channels in the combined optical potential of both lattices.
Emergent phenomena in twisted bi-layers, like the occurrence of superconductivity in twisted graphene~\cite{Cao2018a}, has been observed in fundamentally different physical systems including:
the appearance of quasi-one-dimensional channels along which Abriskosov vortices can freely flow in twisted pinning lattices~\cite{PhysRevB.104.024504},
the formation of flat bands in twisted acoustic metamaterials~\cite{Gardezi2021}, and
enhanced colloidal transport in twisted magnetic patterns~\cite{Stuhlmueller2022}. 
We show here that emergent phenomena arises also in twisted lattices of optical tweezers.

\begin{figure}
  \centering
  \includegraphics[width=1.00\linewidth]{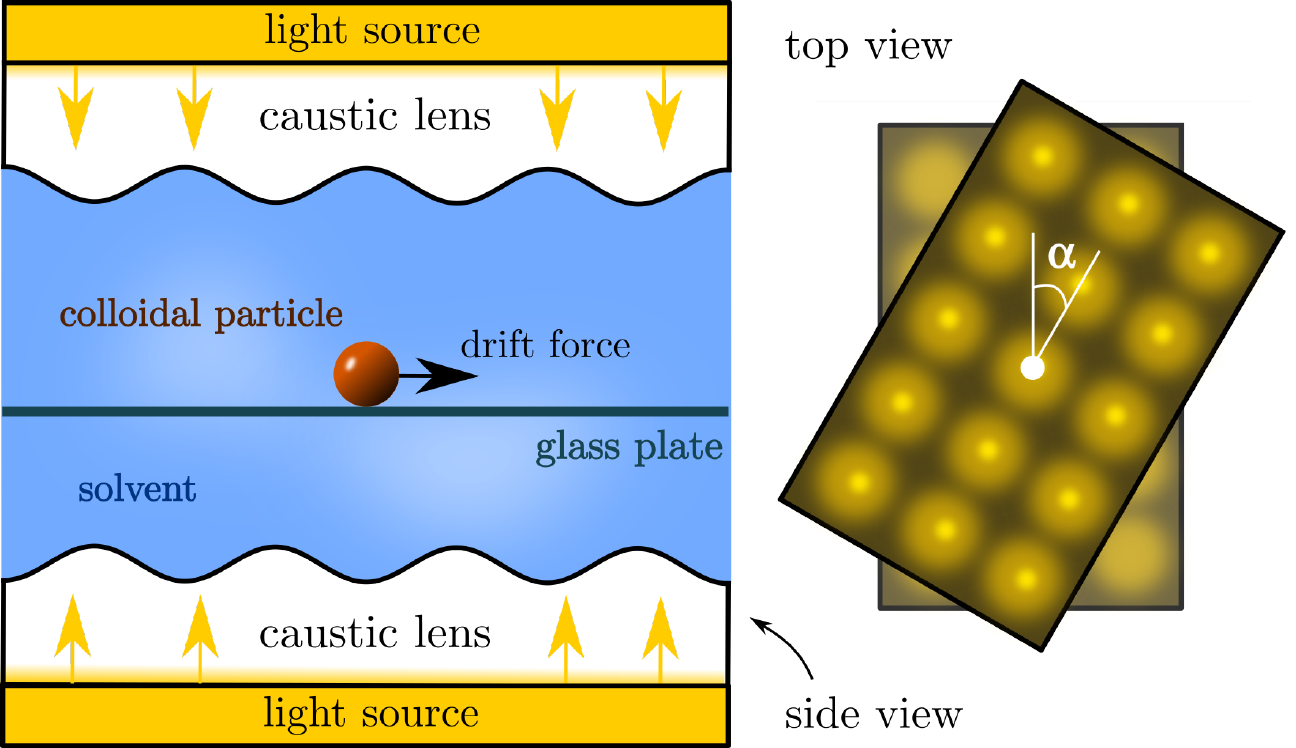}%
	\caption{Schematic of the model: side (left) and top (right) views.
    A colloidal particle immersed in a solvent is located above a glass plate in the middle plane between two lattices of optical tweezers. 
    The lattices are identical but are rotated by an angle $\alpha$ around an axis normal to them.
    The interference between the optical potential of both lattices  creates an anisotropic potential landscape for the colloidal particles.}
\label{fig1}
\end{figure}

\section{Model and Results}

A schematic of the model is shown in Fig.~\ref{fig1}.
The colloidal particles, which are driven by a drift force, are restricted to move in the middle plane between two parallel lattices of optical tweezers.
We consider two periodic lattices of optical tweezers with square and hexagonal symmetries.
The lattices are twisted by an angle $\alpha$ and shifted by half a unit lattice vector.
A destructive interference between the optical potential generated by both lattices results in the formation of channels along which the potential is almost flat.
Using a weak drift force it is then possible to transport the colloidal particles along the flat channels.
At specific twist angles, known as magic angles, the flat channels percolate the entire system allowing transport over arbitrarily long distances.

\begin{figure*}
\includegraphics[width=1.0\linewidth]{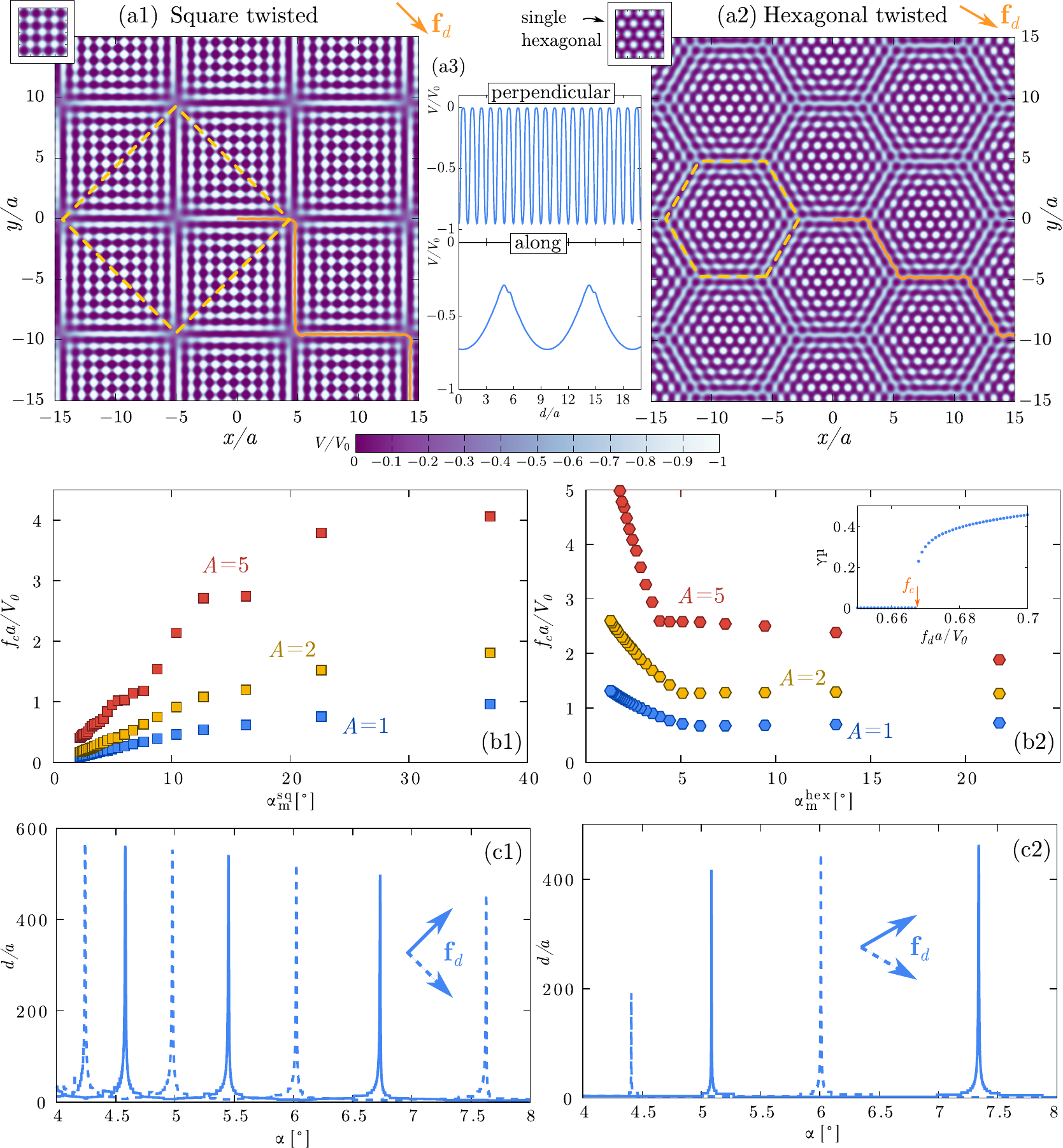}
	\caption{Optical potential (A=1) generated by two square (a1) and two hexagonal (a2) twisted lattices of optical tweezers twisted at magic angles: $\alpha_m\approx6.026^\circ$ in (a1) and $\alpha_m\approx6.009^\circ$ in (a2).
	A super unit cell of the moiré pattern is highlighted (yellow-dashed line).
	A drift force $\f$  points in the direction of two consecutive flat channels (orange arrows) and drives the motion of the particles.
	Characteristic particle trajectories are depicted in orange. The amplitude of the external force is set to $f_d=0.4V_0/a$ (a1) and to $f_d=0.8V_0/a$ (a2).
	The optical potential of single square and hexagonal lattices is shown in the top-left corner of the panels.
	Cuts of the potential (square lattices) perpendicular and along the flat channel that passes through the origin are depicted in (a3) as a function of the distance traveled by a particle initially at the origin.
	The force required to travel along the channel (negative gradient of the potential) is significantly smaller than the force required to travel perpendicular to the channel.
 	Critical force in twisted square (b1) and hexagonal (b2) lattices as a function of the magic angle for different values of $A$.
	The inset in (b2) depicts the mobility $\mu$ vs. the magnitude of the drift force $f_d$ for the optical potential depicted in panel (a1). The orange arrow indicates the value of the critical force $f_c$.
	Distance travelled by a particle originally at the origin in twisted square (c1) and hexagonal (c2) lattices (A=1) as a function of the angle. 
	In (c1) the drift force $f_d=0.4V_0/a$ acts for a total time $t=3000$. The arrows indicate the direction of the drift force, rotated $45\degree$ (solid line) or $-45\degree$ (dashed line)
	with respect to the direction of the first lattice vector.
	In (c2) $f_d=0.8V_0/a$, $t=1000\tau$ and the drift force is rotated $30\degree$ (solid line) or $-30\degree$ (dashed line) with respect to the direction of the first lattice vector.}
	\label{fig2}
\end{figure*}

\noindent{\bf Optical potential.} Following Lacasta et\ al.~\cite{Lindenberg}, we approximate the optical potential at position $\vec{r}$ by
\begin{equation}
  \label{eq:poth}
  V(\vec{r}) = -\frac{V_0}{1+e^{-A(g(\vec{r})-1)}},
\end{equation}
whit positive constants $A$ and $V_0$ that control the steepness and the depth of the optical potential, respectively.
The spatial modulation and the interference between the two arrays of optical tweezers is controlled by the function 
\begin{eqnarray}
  \label{eq:g}
	g(\vec{r}) = \sum_{i=1}^{N}&& \left[ \cos \left(  \qi \cdot \left(\mat{R}_{-\alpha/2}\cdot \r - \frac{\vec{a}_1}{2}  \right)  \right) + \right.\nonumber\\
	&&\left. \cos \left( \qi\cdot\mat{R}_{\alpha/2}\cdot \r \right) \right],
\end{eqnarray}
where each of the two terms in the summation represents one of the lattices of optical tweezers.

The lattices are shifted relative to each other by half of the first lattice vector of the single lattice prior to being rotated, $\vec{a}_1/2$, which we set along the $x-$axis, i.e. $\vec{a}_1=a\vec{\hat{e}}_x$ with $a$ the magnitude of all the lattice vectors.
The relative shift by half of a lattice vector maximizes the destructive interference between the two lattices at the flat channels. 
The matrix $\mat{R}_{\theta}$ is a rotation matrix by an angle $\theta$ around the axis normal to lattice that passes through the origin.
In Eq.~\eqref{eq:g} we rotate each lattice by an angle $\alpha/2$ in opposite directions such that the total rotation between the lattices is the twist angle $\alpha$.
The reciprocal lattice vectors $\qi$ are given by
\begin{equation}
  \label{eq:qi}
	\qi = q \begin{pmatrix}\sin \left(\pi i/N \right)\\ \cos \left(\pi i/N \right)\end{pmatrix},
\end{equation}
where in the square lattice $N=2$ and $q=2\pi/a$, and in the hexagonal lattice $N=3$ and $q=2\pi/ \left( a \sin (\pi/3) \right)$. 

\noindent{\bf Magic angles.}
The total optical potential that results from the interference between both lattices is a moiré pattern.
For specific twist angles the resulting potential is periodic. 
Among those angles for which the potential is periodic, we find the so-known magic-angles, with particularly small lattice constants, given by~\cite{Stuhlmueller2022}
\begin{eqnarray}
  \label{eq:magangles}
  \alpha_m(k,N) = 2\arctan\left(\frac{\frac{1}{Nk+1}\sin(\frac\pi N)}{1+\frac{1}{Nk+1}\cos(\frac\pi N)}\right)
\end{eqnarray}
where $k$ is a natural number and again $N=2$ for the square patterns and $N=3$ for the hexagonal patterns.

The optical potential of lattices twisted at magic-angles, see Fig.~\ref{fig2}(a1) and Fig.~\ref{fig2}(a2), develops super unit cells of length given by approx. $a/(2\sin(\alpha_m/2))$.
That is, the super unit cells grow by decreasing the magic angle.
The super unit cells contain regions where the interference between the lattices is positive and hence the potential resembles that created by a single lattice, shown also as insets in panels (a1) and (a2) of Fig.~\ref{fig2}.
In addition, the super unit cells also contain regions at which the interference is mostly destructive.
There, the potential develops flat channels along which transport is possible using a weak drift force.
The flat channels cross the super unit cell in square twisted patterns, Fig.~\ref{fig2}(a1), and are located at the edges of the super unit cells in hexagonal twisted patterns, Fig.~\ref{fig2}(a2).
We show in Fig.~\ref{fig2}(a3) a cut of the optical potential in the directions perpendicular to a flat channel and also along the flat channel, as indicated.
The force required to travel along the flat channel (given by the negative gradient of the optical potential) is significantly weaker than that require to travel perpendicular to the flat channel.
Increasing the parameter $A$ makes the potential flatter along the central region of a flat channel.
However, it also makes the potential steeper near the intersections between two flat channels.
At the magic angles the flat channels percolate the entire system. 

\noindent{\bf Computer simulations.} We neglect inertial effects and therefore use overdamped dynamics to simulate the motion of a single colloidal particle.
At high laser intensity the Brownian forces can be neglected as compared to the optical forces, we therefore set the temperature to zero such that Brownian motion does not hinder the phenomenology.
The equation of motion for a single particle reads:
\begin{equation}
  \label{eq:bd}
 \gamma \dot{\r} = -\nabla V(\r) + \f
\end{equation}
where $\gamma$ is the friction coefficient against the implicit solvent, $\dot{\r}$ indicates the time derivative of the position vector, and $\f$ is an homogeneous external drift force.
The magnitude of a lattice vector $a$, the energy parameter of the optical potential $V_0$, and the friction coefficient $\gamma$ define our system of units. 
The intrinsic time-scale is therefore $\tau=\gamma a^2/V_0$. We integrate the equation of motion using an adaptive Heun-Euler-scheme~\cite{adaptiveBD},
setting the relative allowed error per time-step to $10^{-2}$ and the absolute allowed error in the positions to $10^{-4}a$.

\noindent{\bf Drift force.}
To drive the colloidal motion we use a drift force $\f$ pointing along the bisector of the directions of two flat channels, see Figs.~\ref{fig2}(a1) and~\ref{fig2}(a2).
Hence,
\begin{equation}
	\f=f_d \begin{pmatrix} \cos\alpha_d \\ \sin\alpha_d\end{pmatrix},
		\label{eq:drift}
\end{equation}
where the angle is $\alpha_d(k)= (-1)^{k}{\pi}/{4}$ in square lattices and $\alpha_d(k)=(-1)^{k}{\pi}/{6}$ in hexagonal lattices, and the index $k\in\mathbb{N}$ is the same as for the magic angles in eq.~\eqref{eq:magangles}.
The prefactor $(-1)^{k}$ alternates the direction of the drift force  between the first and the fourth quadrants, reflecting the fact that the flat channels that support transport alternate from one magic angle to the next one.

Panels (b1) and (b2) of Fig.~\ref{fig2} show the magnitude of the critical drift force $f_c$ required to transport colloidal particles along the flat channels.
To calculate $f_c$ we measure in the simulations the colloidal mobility $\mu$ under the influence of the drift force:
\begin{equation}
	\mu = \frac{\left| \Delta\vec{r}(t_f) \right|}{t_ff_d},
\end{equation}
where $\Delta\vec{r}(t_f)$ is the distance traveled by a particle during a total time $t_f=3000\tau$ in twisted square lattices, and $t_f=1000\tau$ in twisted hexagonal lattices.
The colloidal mobility vanishes for weak drift forces, increases rapidly at the critical drift force $f_c$, and it saturates for strong drift forces, see an example in the inset of Fig.~\ref{fig2}(b2).

In square lattices, the critical force at which transport along the channels is activated decreases monotonically by decreasing the magic angle, see Fig.~\ref{fig2}(b1).
In hexagonal lattices, Fig.~\ref{fig2}(b2), the critical force presents two distinct regimes.
First, for small magic angles, there is a rapid decrease of $f_c$ by increasing the magic angle.
In the second regime, depending on the steepness of the potential, the critical force either slightly increases [e.g. $A=1$ in Fig.~\ref{fig2}(b2)] or it slightly decreases [e.g. $A=5$ in Fig.~\ref{fig2}(b2)].
For both square and hexagonal lattices, increasing the steepness of the potential $A$ also increases the magnitude of the critical drift force.
Increasing $A$ makes the potential flatter along the channels, but it also increases the steepness of the potential at the intersection between two flat channels which results in higher values of the magnitude of the critical force.

In Fig.~\ref{fig2}(c1) and Fig.~\ref{fig2}(c2) we represent the distance travelled by a particle, $d$, as a function of the twist angle for both square and hexagonal lattices, respectively.
The particle is at time zero located at the axis of rotation of both lattices (i.e. in the middle of a flat channel).
The motion is driven by a drift force acting for a total time $3000\tau$ ($1000\tau$) in square (hexagonal) lattices, and whose magnitude is larger than the critical force required to move particles at any of the magic angles that occur in the represented range of twist angles.
For each lattice, we plot two curves, corresponding to drift forces that according to Eq.~\eqref{eq:drift} point either in the first or in the fourth quadrant.
The curves clearly show that the edges that support transport alternate from one magic angle to the next one. 
The distance travelled by the particles present sharp peaks at the magic angles and hence even a small deviation from the magic angle has a marked effect on the transport.
In square lattices the value of $d$ at the magic angles decreases by increasing the magic angle since the critical force increases with the magic angle, see Fig.~\ref{fig2}(b1), and we keep the magnitude of 
the drift force constant.
The opposite behaviour is observed in hexagonal lattices for the range of angles shown in Fig.~\ref{fig2}(c2). That is, $d$ at the magic angles increases by increasing the magic angle.
For the range of angles shown in Fig.~\ref{fig2}(c2), the critical drift force in hexagonal lattices decreases by increasing the magic angle which explains the observed travelled distance at the magic angles.
\section{Conclusions}

Despite being substantially different systems, the colloidal transport in twisted optical lattices is quite similar to the transport of magnetic colloidal particles in twisted magnetic patterns~\cite{Stuhlmueller2022}.
The reason is that the magnetic potential in twisted magnetic patterns corresponds to the limit $A\rightarrow0$ of the optical potential in twisted lattices of optical tweezers.
The setup described here offers therefore more flexibility since the steepness of the potential of single lattices, controlled by the parameter $A$, can be adjusted experimentally by varying the width of the tweezers.
In addition, optical tweezers offer the advantage that magnetic but also non-magnetic colloidal particles can be used in the experiments.

Several types of external forces are available experimentally to drive the motion~\cite{Loewen2008}.
These include among others electric and magnetic fields, pressure gradients, and the gravitational field of Earth in the case of micron-sized colloidal particles with a substantial contrast between the bare and the solvent mass densities.

There exist other twist angles for which the combined potential of the two lattices is also periodic~\cite{Stuhlmueller2022}. However, for those angles, the direction along which transport is possible along the flat channels changes inside the super unit cell. Hence, stronger drift forces are required to cause macroscopic transport.

We have focused here on the dilute regime where interparticle interactions do not play any role. 
Interesting collective effects appear in many-body particle systems driven on periodic landscapes, including structural transitions and directional locking~\cite{PhysRevLett.106.060603,AReichhardt2016,AReichhardt2012}.
It would be also interesting to study collective effects in the dynamics of many-body particles in twisted lattices such as the superadiabatic forces~\cite{PhysRevLett.125.018001,RevModPhys.94.015007} and the occurrence of solitons~\cite{https://doi.org/10.48550/arxiv.2204.14181}.

Another interesting extension of the present work is the characterization of the transport in twisted three-dimensional optical lattices.
Moreover, using periodic two-dimensional magnetic patterns together with a homogeneous magnetic field, one can topologically transport magnetic colloidal particles placed above the patterns~\cite{Loehr2016,Heras2016,Loehr2017}.
There exist special modulation loops of the orientation of the external field such that once the loop returns to its initial position the particle has been transported by one unit cell above the pattern.
The colloidal motion is topologically protected and takes place in a plane due to the two-dimensional nature of the magnetic patterns.
Optical potentials could be used to extend the study of topologically protected colloidal transport to three-dimensional systems.

\section{Acknowledgements}
This work is funded by the Deutsche Forschungsgemeinschaft (DFG, German Research Foundation) under project number 440764520.

\end{document}